# From Cathode to Anode: Understanding Lithium Loss in 21700-type Cells with Ni-rich NCM Cathodes and Graphite/ Silicon Oxide Anodes

Thien An Pham[a,b], Hannah Bosch[c], Giovanni Ceccio[d], Lukas Keller[e], Hannes Wolf[f], Nicolas Bucher[g], Peter Müller-Buschbaum[b], Ralph Gilles[a]

## Affiliations

*[a]* *T. A. Pham, R. Gilles*
*Technical University of Munich*
*Heinz Maier-Leibnitz Zentrum*
*Lichtenbergstraße 1, 85748 Garching, Germany*

*[b]* *T. A. Pham, P. Müller-Buschbaum*
*Technical University of Munich*
*TUM School of Natural Sciences, Department of Physics, Chair for Functional Materials*
*James-Franck-Str. 1, 85748 Garching, Germany*

*[c]* *H. Bosch*
*Technical University of Munich*
*TUM School of Engineering and Design, Department of Energy and Process Engineering, Chair for Electrical Energy Storage Technology*
*Arcisstraße 21, 80333 Munich, Germany*

*[d]* *G. Ceccio*
*Academy of Sciences of the Czech Republic*
*Nuclear Physics Institute*
*CZ-25068 Řež, Czech Republic*

*[e]* *L. Keller*
*Paul Scherrer Institut PSI*
*Forschungsstrasse 111, 5232 Villigen PSI, Switzerland*

*[f]* *H. Wolf*
*BASF SE*
*Carl-Bosch-Str.38, 67056 Ludwigshafen, Germany*

*[g]* *N. Bucher*
*VARTA Microbattery GmbH*
*VARTA-Platz 1, 73479 Ellwangen, Germany*

# Abstract

Moving to larger cell formats in batteries increases the overall useable energy. However, inhomogeneities can emerge in larger cell formats that influence the aging behaviour. This study investigates the degradation of NCM cathodes and graphite/SiOx anodes in 21700-type cells with cyclic aging using *in operando* neutron diffraction and neutron depth profiling. Prolonged cycling resulted in Li loss, evident on the cathode side from the reduced NCM unit cell expansion during charging. On the anode side, this Li loss was observed as a diminished formation of the fully lithiated $LiC_6$ phase over extended cycling. From the differential voltage analysis over the course of cyclic aging, the loss of active anode material was detected in addition to the loss of Li inventory. This was confirmed by the diffraction data as the transition to $LiC_{12}$ and the $LiC_6$ formation onset shifted to lower capacity values, indicating active material loss, as less Li was needed to trigger the transitions. Li concentration profiles across different electrode positions supported the Li depletion in the cathode, while an elevated Li concentration in the anode suggested increased solid-electrolyte interphase formation. This finding suggests that Li from the cathode is consumed on the anode side. Notably, samples from the cell center exhibited more severe degradation than those from the bottom edge, highlighting existing inhomogeneities along the height of the cell.

## Idea for TOC Figure

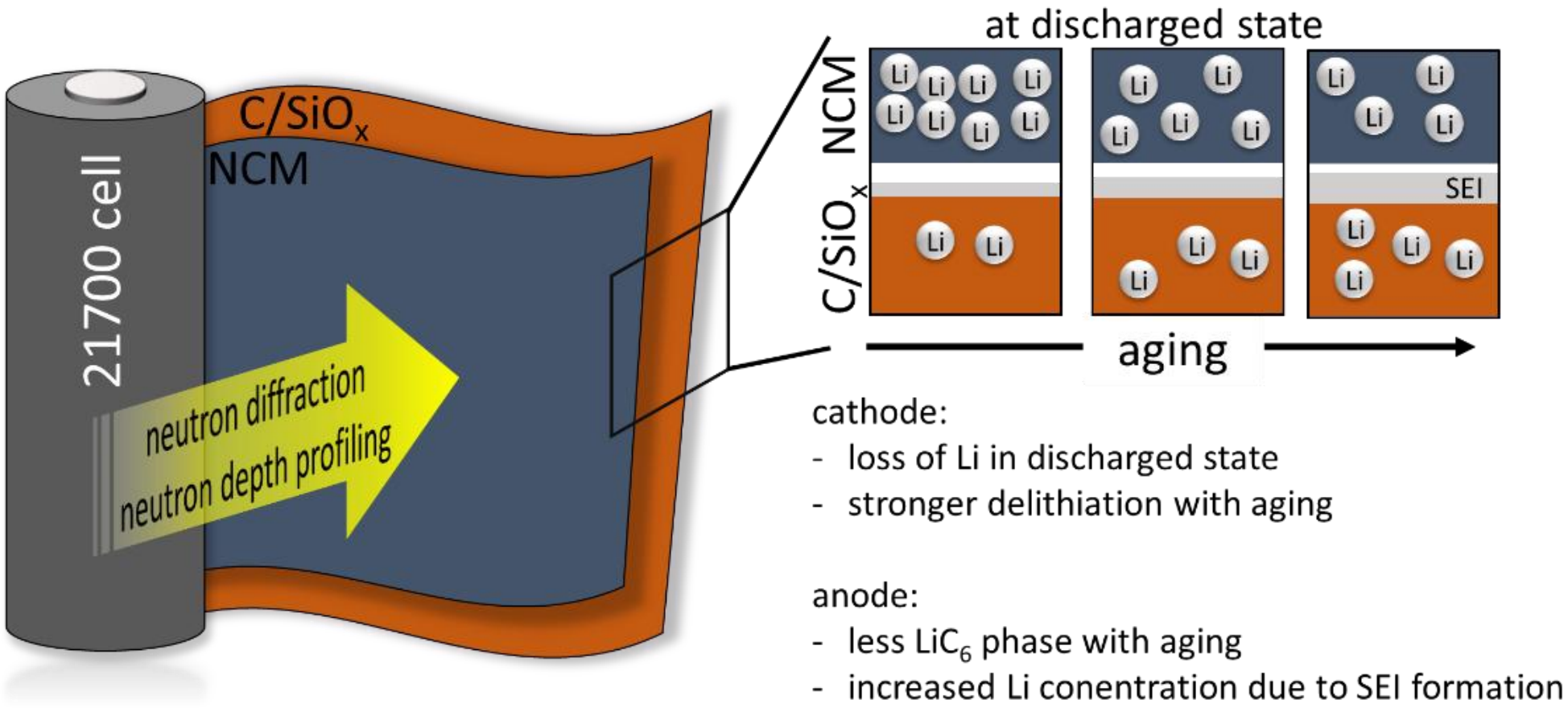

# Introduction

In recent years, lithium-ion batteries (LIBs) have become the dominant energy storage solution across various applications, ranging from consumer electronics to the transportation sector. The ongoing transition from combustion-engine vehicles to electromobility plays a critical role in reducing greenhouse gas emissions. As a result, extensive research has been dedicated to improve electric vehicle (EV) battery technology to enhance energy density, efficiency, and longevity.

LIBs for EVs have evolved in both chemistry and cell format over time. The three primary cell designs used in the industry today are cylindrical, prismatic, and pouch cells. Among them, cylindrical cells and particularly the 18650 format have been widely adopted due to their robust mechanical stability and well-established manufacturing processes. More recently, larger cylindrical formats such as the 21700-type and even 4680-type cells have been introduced to increase energy while maintaining similar production costs. [1,2] However, scaling up cell size can introduce inhomogeneities, such as thermal gradients, which can affect the performance and cycle life, and mechanical stresses. [3–5] These inhomogeneities are reflected in the lithiation behaviour of the electrodes. [6,7] Petz et. al detected intensified degradation of graphite anodes in the middle position of 21700-type cylindrical cells. [8] Lately, it was reported that the spatial orientation of cylindrical cells as well as the electrolyte amount lead to variations in ion concentrations due to electrolyte movement (electrolyte motion induced salt inhomogeneities, EMSI), which causes increased capacity fading. [9–11]

Despite their advantages, LIBs undergo degradation over time, affecting the electrodes. A key focus of battery research has been to understand the aging mechanisms in the active materials of LIBs. For cathodes, Ni-rich layered transition metal oxides such as $LiNi_xCo_yMn_{1-x-y}O_2$ (NCM) and $LiNi_xCo_yAl_{1-x-y}O_2$ (NCA) are widely employed to maximize energy density. However, these materials are susceptible to irreversible surface reconstruction from the layered structure to the rock-salt phase, induced by high voltages and prolonged cycling. This can ultimately release oxygen, leading to further degradation pathways such as formation of hydrofluoric acid (HF) and electrolyte decomposition. [12–15] Transition metal (TM) dissolution has also been reported as a degradation mechanism. [16] Here, the TM ions migrate to the anode, damaging the solid-electrolyte interphase (SEI). Particularly, Mn dissolution has been identified as a critical factor in graphite anode degradation, as $Mn^{2+}$ ions destabilize the SEI and increase lithium consumption during reformation, leading to capacity loss. [17]

On the anode side, graphite (C) remains the most commonly used anode material due to its well-understood intercalation mechanism and stable cycling performance. [18] However, alternative anode materials, such as silicon (Si) and silicon oxides ($SiO_x$), have been explored to enhance energy density. [19] Si-based anodes offer higher theoretical capacity than graphite but suffer from large volume expansion during lithiation, which promotes crack formation and SEI instability, causing its continuous reformation. The SEI formation is considered as the cause for the loss of Li inventory (LLI). To mitigate these issues, graphite-silicon composite anodes have been developed, utilizing the high energy density of silicon while minimizing excessive expansion effects. [20–22]

Aging behavior in cylindrical LIBs has been extensively studied, but detailed insights into the degradation of larger cylindrical cells remain an area of active research. In this work, the effects of aging on 21700-type cylindrical cells with NCM cathodes and graphite/$SiO_x$ composite anodes were investigated to identify key degradation mechanisms. To achieve this, 21700-type prototype cells were cycled to different states of health (SOH) and compared to an uncycled reference cell. To track lithium transport and phase transitions within the electrodes, *in operando* neutron diffraction was employed and the results were combined with electrochemical data to gain deeper insights into lithium loss mechanisms. Additionally, neutron depth profiling (NDP) measurements were performed on the extracted electrodes to determine the Li concentration as a function of depth for two different SOH to validate the loss of Li. This study provides a comprehensive understanding of the interplay between electrode degradation and lithium loss in high-energy-density cylindrical cells, contributing to the optimization of next-generation LIBs for EV applications.

# Experimental

## Battery Samples

The 21700-type cylindrical cells were manufactured by VARTA, Germany and are prototype LIBs and consisted of of a $LiNi_{0.83}Co_{0.12}Mn_{0.05}O_2$ cathode (NCM-831205, supplied by BASF SE, Germany, active material content (AM) 96.5%) cathode and an anode blend (95% AM) of graphite with approximately 14 wt.% of $SiO_x$. The NCM-831205 cathode material was supplied by BASF SE. The theoretical capacity of the electrodes were 190 mAh/g (192 g/m²) and 510 mAh/g (76 g/m²) at an electrode density of 3.35 g/cm³ and 1.6 g/cm³ for the cathode and anode, respectively. A PAA/SBR-binder system was used for the anode. The cathode binder is PVDF. The electrolyte contained 1.15 M LiPF6 in DMC:FEC:EMC 70:20:10 vol.%. A 10 µm copper and 15 µm aluminium foil was used as current collector, respectively and a 12 µm PE/ceramic separator was built into the cells.

## Electrochemical Tests

All cells were positioned in a climate chamber and cycled at 25 °C within the voltage range of 2.5 to 4.2 V using a CTS Standard from Basytec.

Before starting the cyclic aging protocol, the cells underwent an initial Check Up procedure, with two cycles of C/3 constant current (CC) constant voltage (CV) charge (CH) and CC discharge (DCH), to determine the capacity of each individual cell. Direct current internal resistance (DCIR) measurements (10 s pulse with C-rate of C/3, 2/3C and 1C at a 50% SOC) were also part of the initial Check Up procedure. Additionally, the pOCV curve was determined with C/20 $CC_{Ch}$ | $CC_{DCH}$.

One cell was chosen for each data point. The SOH100 cell was stored at 25 °C whilst the SOH80 and SOH70 cell were cycled according to the following protocol until they reached an SOH of either 80% or 70%. After two stabilisation cycles of C/10 $CCCV_{CH}$ | $CC_{DCH}$ the cells underwent the cyclic Check Up procedure. This consisted of two cycles of C/3 $CCCV_{CH}$ | $CCCV_{DCH}$, followed by two cycles of C/10 $CCCV_{CH}$ | $CCCV_{DCH}$ and DCIR measurements (10 s pulse with C-rate of C/3, 2/3C and 1C at a pre-defined SOC). The cyclic Check Up was repeated after two cycles of C/2 $CCCV_{CH}$ | 2C $CC_{DCH}$ and the 45 subsequent cycles of C/2 $CCCV_{CH}$ | $CC_{DCH}$ (Table 1). The cut-off current of all CV phases was set to C/50.

*Table 1: Full testing protocol of the 21700-type cylindrical cells.*

<table>
<tr><th></th><th></th><th>Cycle Information</th><th>C-Rate</th></tr>
<tr><td></td><td>Stabilisation Cycles</td><td>2 cycles</td><td>C/10 $CCCV_{CH}$ | $CC_{DCH}$</td></tr>
<tr><td rowspan="5">Repeat until SOH of 80% or 70%</td><td>Check Up C/3</td><td>2 cycles</td><td>C/3 $CCCV_{CH}$ | $CCCV_{DCH}$</td></tr>
<tr><td>Check Up C/10</td><td>2 cycles</td><td>C/10 $CCCV_{CH}$ | $CCCV_{DCH}$</td></tr>
<tr><td>DCIR</td><td>10 s pulse</td><td>C/3, 2/3C, 1C</td></tr>
<tr><td>Pre-Step</td><td>2 cycles</td><td>C/2 $CCCV_{CH}$ | 2C $CC_{DCH}$</td></tr>
<tr><td>Cycling</td><td>45 cycles</td><td>C/2 $CCCV_{CH}$ | $CC_{DCH}$</td></tr>
</table>

The capacity values were taken from the second cycle of the discharge step of the cyclic Check Up. The SOH was calculated by dividing the discharge capacity (either C/3 or C/10) with the discharge capacity of the first cyclic Check Up (either C/3 or C/10). For the cell SOH70, the cycling process was stopped prematurely before reaching the termination criterion of an SOH of 70% because of the impending *in operando* neutron diffraction experiment.

For the incremental capacity analysis (ICA) and differential voltage analysis (DVA), the second cycle of the cyclic Check Up at C/10 was used. A Savitzky-Golay finite impulse response filter was applied in Matlab after the derivation to smooth the data points. [23,24] The DCIR data is not shown, as the correct SOC was not reached over cyclic aging and thus, making DCIR analysis over aging not feasible.

## *In Operando* Neutron Diffraction

*In operando* neutron diffraction experiments were performed on the DMC instrument at the SINQ, Paul Scherrer Institute (PSI) in Villigen, Switzerland. [25] The identical cylindrical cells at SOH70 and SOH80 were measured and compared to an uncycled cell at SOH100. The cells were mounted on a 3D-printed sample holder and connected to the sample stage via a solid Al rod. The electrical contacts were soldered to the top of the cell housing and connected to the BioLogic VSP-300 with a 10 A booster. The cell was first discharged using a current rate of C/10 with constant current (CC) to 2.5 V and subsequently charged to 4.2 V with the same current. The SOH100 cell was the only cell that included a CV phase after the CC charge. In order to keep the comparability, only the CC charge steps were compared in the analysis. The cell was aligned so that the center of the cell was measured with ND.

Powder patterns were collected by measuring at six detector positions for 30 seconds with a slight shift in $2\theta$ to minimize the oscillations of the radial collimator. These six measurements were then summarized to create the whole pattern, resulting in an acquisition time for one pattern of 3 min (Figure 1b).

For the analysis of the *in operando* ND, GSAS-II [26] and Matlab were used. The lattice parameters of NCA were determined with Rietveld refinement using a model based on the NCM structure that was measured with X-ray diffraction on the pristine active material. [paper Jonas & Lukas] The background was determined using the automatic background function of GSAS-II. First, the sample displacement was obtained using the Fe phase, which stems from the cell housing since it has the strongest contribution to the diffraction pattern. The Cu phase from the anode current collector defined the layer displacements since their reflection peaks stayed constant during the experiment. The value was then transferred to the NCM phase. A reference consisting of $Na_2Ca_3Al_2F_{14}$, $CaF_2$ and $Na_3AlF_6$ was measured in the same instrument setup to determine the instrumental broadening and the wavelength of 2.454 Å.

The graphite anode was analyzed by performing single-peak fitting in Matlab using a pseudo-Voigt function. For the transition from C to $LiC_{12}$, the peak height and position were fitted in the $2\theta$-range of 39° to 45°. For the $LiC_6$ reflection, the position was kept constant while the intensity was fitted.

## Neutron Depth Profiling

Neutron depth profiling (NDP) was performed on the TNDP spectrometer of the NPI CANAM infrastructure at the nuclear research reactor LVR15 in Rez. The NDP is a non-destructive technique, allowing the depth-dependent detection of Li by the nuclear reaction between neutrons and Li atoms. $LiNbO_3$ was used as a reference sample to calibrate the energy and elemental abundance. The $LiNbO_3$ reference sample contains 1.7e22 Li/cm$^3$ and was cross-checked using an SRM2137 boron standard from NIST. The data was normalized using the fluence of the neutron beam during the experiment.

After the *in operando* neutron diffraction experiments, the cylindrical were discharged and opened to extract the electrodes at different positions. The samples were punched out using a coin cell punching tool with a diameter of 14 mm and washed with dimethyl carbonate to remove excess Li-salt and electrolyte. Samples were taken from both cathodes and anodes at different positions: at the centre in the middle and the bottom and outside in the middle (Figure 1c). Each sample was sandwiched between a 7.5 µm Kapton foil and a 100 µm Kapton foil with adhesive and was separately sealed in Al pouch bags with argon bags for transportation and storage until the measurements (Figure 1d). During the measurement, the $^4He$ particles were absorbed by the Kapton foil, and only the signal from the more energetic $^3H$ particles was analyzed.

The N4DP software was used for the analysis. [27] The analysis procedure is based on our previous studies. [28,29] To calculate the depth profile, uniform densities were assumed for the cathode of $\varrho_{\mathrm{NCM}} = 3.35\ \mathrm{g/cm^2}$ and anode $\varrho_{\mathrm{C/SiOx}} = 1.6\ \mathrm{g/cm^2}$ that were provided by the cell manufacturer for the pristine electrodes.

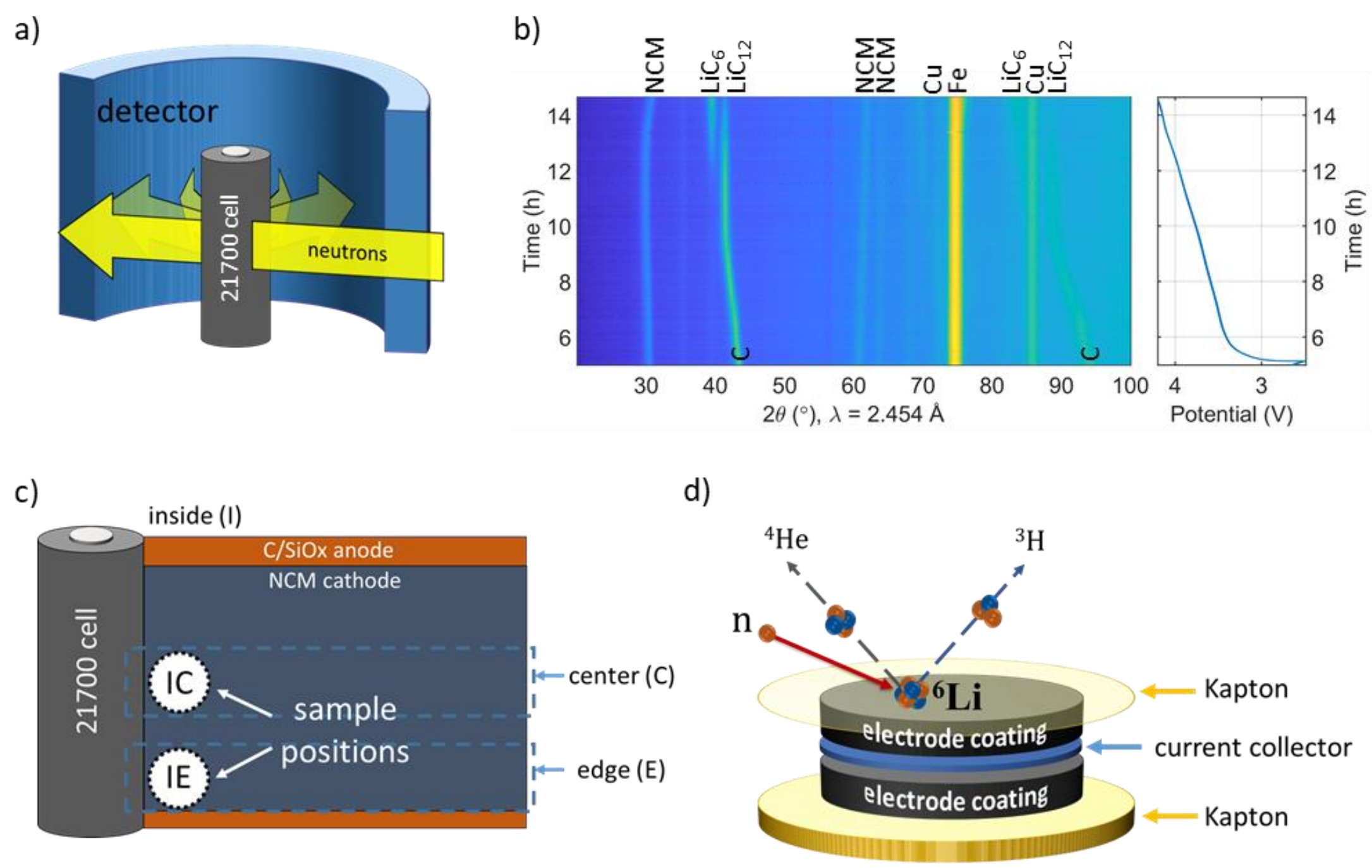

*Figure 1: Experimental setup for the in operando ND experiment (a) and the 3d plot of the collected diffraction patterns for the SOH100 cell during the charge from 2.5 to 4.2 V at C/10 (b). Positions on the electrode where the samples were harvested for NDP experiment (c). The electrodes were put in between Kapton foil, which absorbed the $^4$He particles (d).*

## Results and Discussions

### Electrochemical Cycling

#### Capacity Decrease

The discharge capacity at different C-rates from the cyclic aging procedure is illustrated in Figure 2 for the SOH80 and SOH70 cells. The decrease in discharge capacity of both cells is in good agreement, and the capacity fade is consistent for all C-rates, which emphasizes the high comparability of these cells. Cycles at a lower current of C/10 and C/3 reach a smaller capacity than at C/2 while maintaining the same slope of the capacity fade. Increasing the current to 3C/2 yields lower discharge capacities as the higher currents lead to a more substantial potential drop, and thus, the cut-off voltages are reached earlier. The overall capacity decrease at C/2 is approximately linear. However, at the beginning of every set of C/2 cycles, the first discharge capacity value is higher than the last value of the previous set, indicating that some of the lost capacity has been regained during the C/10 check-up cycles. [30]

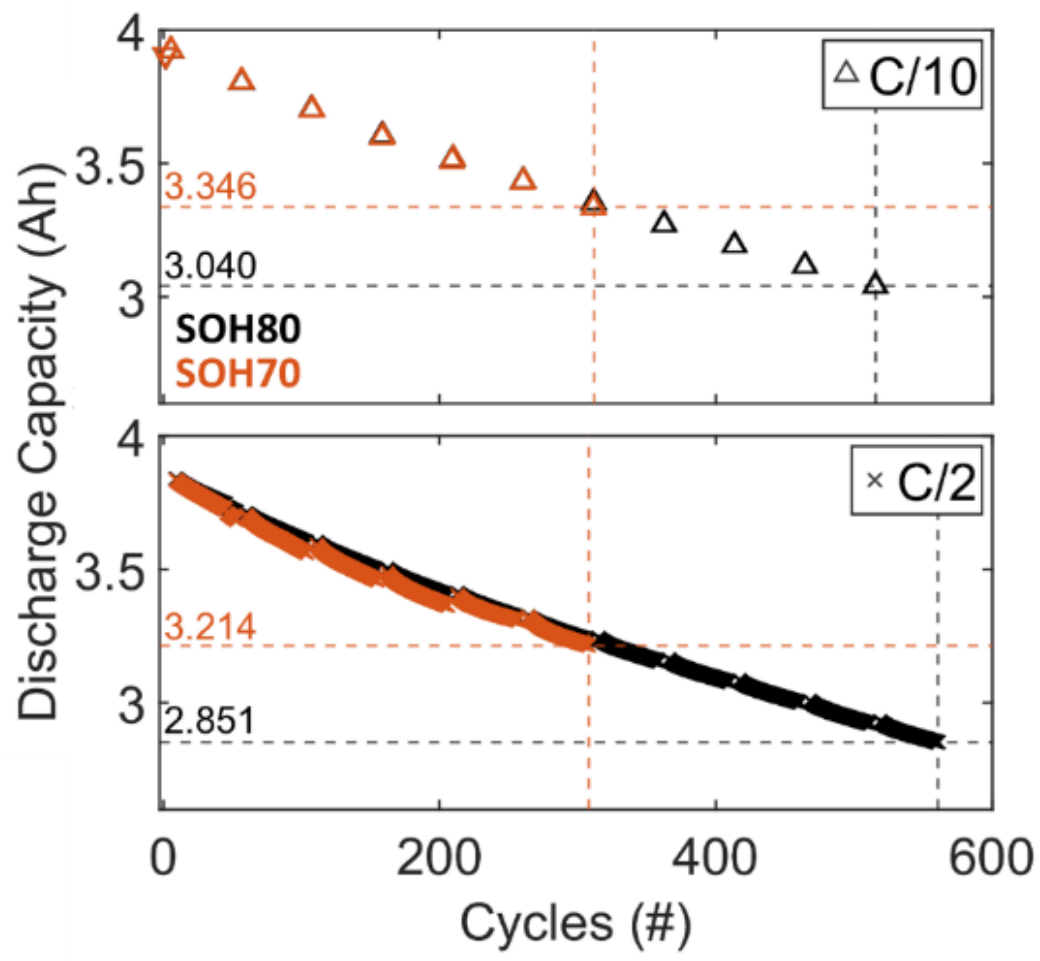

*Figure 2: Discharge capacities during the cyclic aging for the SOH80 (red) and SOH70 cells (black) for different C/10 and C/2.*

#### ICA/DVA Analysis

The check-up cycles at C/10 are shown in Figure 3a for the SOH70 cell. With aging, higher potentials are required to reach the same state of charge (SOC), indicating the build-up of internal resistance and, thus, overpotentials. The DVA curves are calculated from the check-up cycles at C/10, and the capacity loss is reflected in the movement of the peaks (Figure 3b). At low SOC, a small peak is visible at begin of life around $Q_{SiOx} = 0.7\mathrm{Ah}$, which is assigned to the lithiation of the $SiO_x$. The capacity contribution of $SiO_x$ is calculated using the total charge capacity $Q_{charge}$ and is defined as $q_{SiOx} = Q_{SiOx}/Q_{charge} = 0.181$. With an amount of 14 wt.% of $SiO_x$ in the blend anode, this corresponds to a specific capacity of $Q_{SiOx} = 505\mathrm{mAh/g}$, which lies in the reported range of 400 to 2400 mAh/g for $SiO_x$ depending on the oxygen content. [31–34] With ongoing cycling, the DVA peak of $SiO_x$ slowly moves to lower capacities and almost vanishes, indicating that less $SiO_x$ remains active during cycling.

The DVA peak at high SOC, labelled as MaxHi, is determined for the analysis based on Lewerenz et al. [35]. During charging, lithium from the cathode transfers to the anode and is stored between the graphene layers. MaxHi marks the phase transition in the anode from the $LiC_{12}$ phase to the fully lithiated $LiC_6$ phase, and thus, its position gives insight into the degradation of the anode. The position of MaxHi decreases with cycling (Figure 3c), demonstrating that less capacity is required for the transition to the $LiC_6$ phase and indicating loss of active anode material (LAAM). The decrease is approximately linear, which suggests a steady LAAM. For the discharge, the DVA curve is inversed. Therefore, the transition from $LiC_{12}$ to $LiC_6$ is given by the minimum at high SOC, labelled as MinHi. [35]

The degradation modes are seen more clearly by shifting all the DVA curves for MaxHi or MinHi to overlap for all the curves (Figure 3d & e). In addition to the LAAM, the loss of Li inventory (LLI) is visible, which is defined by the shift of the DVA curves from high SOC to MaxHi and MinHi. [35–37] Thus, less Li is stored in the anode after transitioning to the $LiC_6$ phase. Another way to identify the degradation modes is to analyse the ICA curves. In Figure 3f, the ICA curves during charge and discharge are displayed. The peak intensities decrease with ongoing cycling while the positions change to different potential values except for the strongest peak around 4.15 V. Different points of interest were marked and labelled based on the ICA curves for NCM/C+Si from Dubarry et al. [38]. ①, ②, ③ and ④ mark the ICA peaks from the NCM while ❶, ❷ and ❺ describe the contribution of the graphite and ❷$_{Si}$ represents the silicon in the anode. The degradation is estimated from the movement of these ICA peaks. Since the ICA was performed using the full cell voltage, some contributions from the electrodes overlap and thus, not all peaks present in the half cell data are visible in the ICA. The shift of the ❺★③ and ❷★③ peaks downwards and to higher potentials during charging and the ❶★② shift to lower potentials are an indication for LLI in the cell. The same movement for ❺★③ and ❷★③, while ❶★② stays constant, demonstrates the loss of lithiated graphite intercalation compounds (GIC). Additionally, the right shift of the ❷$_{Si}$★④ reflects the loss of lithiated Si. Generally, the detected changes in the ICA curves result from the combined effects of LLI and LAAM, which agrees with the results from the DVA.

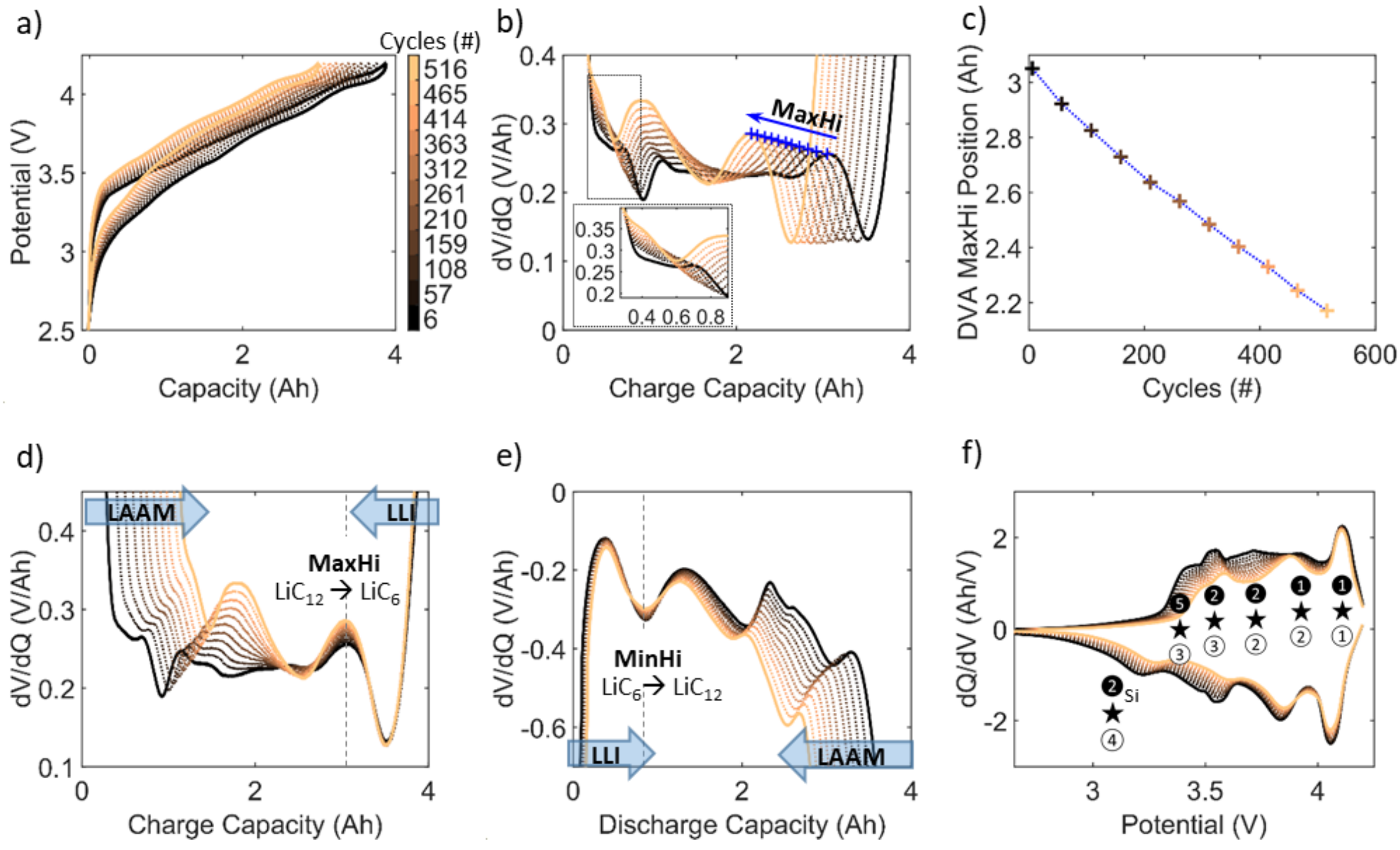

*Figure 3: Analysis of the check-up cycles at C/10 (a) of the SOH70 cell. Differential voltage analysis (dV/dQ) during charge with magnified region for the SiOx peak (b) and the position of the MaxHi peaks over the course of cycling (c). Shifted DVA curves for charge (d) and discharge (e), shifted so that the MaxHi and MinHi peak overlap for all the cycles. ICA curves with marked transitions (f).*

## *In Operando* Neutron Diffraction

### Change of the NCM Unit Cell

Based on the *in operando* ND experiment, the structure changes of the NCM are derived via the shift of the reflection peaks during charging. The position of the NCM 003 and NCM 101 reflections are shown in SI 2. The position of the NCM 003 peak first moves to lower values and increases strongly afterwards. Its movement is strongly correlated to the $c$ parameter, which describes the vertical elongation of the NCM unit cell. The NCM 101 and 113 reflections increase monotonously during charging and contribute to the $a$ parameter. Upon charging, the NCM is delithiated and the transition metals (TM) are oxidized to maintain the charge neutrality, which leads to a stronger attraction between the TM and oxygen atoms and, thus, the contraction of $a$ (SI 3). With decreased amounts of Li, their screening effect of the negative charge of the oxygen atoms is weakened. It causes an increased repulsion between the oxygen slabs, increasing the $c$ parameter. [39–41]

With increased cyclic aging, the movement of the reflection peaks is decreasing, which strongly indicates the loss of Li. The Li loss is prominently visible in the discharged state, reflected in the evolution of the lattice parameters (SI 3). The capacity values were shifted along the x-axis so that the peak visible in the $c/a$ curve overlaps for all the cells (SI 3b & Figure 4a). With ongoing cyclic aging, less Li can be retrieved from the NCM cathode, shifting the starting position to the right towards higher capacity values. This is evident in the lattice parameters at 0% and 100% SOC (Figure 4b), particularly the $a$ parameter, whose nearly linear evolution provides a general reliable capacity estimation. The $a$ parameter at $\mathrm{SOC} = 100\%$ stays almost the same with aging while the value at $\mathrm{SOC} = 0\%$ decreases, meaning that Li is lost in the discharge step. Put differently, Li is consumed on the anode side and, thus, the amount of Li returning from the anode during discharge decreases with aging. To estimate the Li loss, the total movement of the $a$ parameter is determined by calculating the difference between the two values. The relative decrease of the $a$ parameter movement from SOH100 to SOH70 is -34%, showing that the less Li is accessible with further cycling.

Figure 4b illustrates the $c$ values in the charged and discharged states for the three different SOH states. The increase of the value at $\mathrm{SOC} = 0\%$ is comparable to the decrease of the a parameter at $\mathrm{SOC} = 0\%$. At $\mathrm{SOC} = 100\%$, the $c$ parameter reaches lower values for the stronger-aged cells. With prolonged cycling, the electrode potentials shift to higher values while the cell voltage remains the same, leading to the increased delithiation of the cathode. [42–44] This implies that the NCM is delithiated to higher degrees, which in return releases additional cyclable Li. However, the Li loss in the discharged state still outweighs the Li gain. At the same time, the higher cathode potentials promote parasitic electrolyte oxidation, which increases the cell degradation. [12,45] For cells with NCM cathodes, the decomposition of ethylene carbonate (EC) from the electrolyte at high potentials is reported to cause the release of singlet oxygen ($^1O_2$) from the NCM, leading to the formation of $H_2O$. [13,46] The $H_2O$ is reduced on the anode side, forming LiOH and $H_2$, and aggravates the SEI formation. [17,47]

From the lattice parameters of the NCM, the volume of the unit cell can be calculated (Figure 4d). Similarly to the evolution of the $c$ parameter, the unit cell contraction during cycling grows stronger with extended aging even though the transferred amount of Li reduces (Figure 4e). This is caused by the rapid decrease of the $c$ parameter at higher delithiation. Ultimately, a higher volume change can lead to the accumulation of stress in the material, which results in the particle crack formation of the active material. [48,49] Especially, the anisotropic stress accumulation in NCM, which reaches its maximum around $\mathrm{SOC} = 70\%$, strongly contributes to the degradation of the cell. [50]

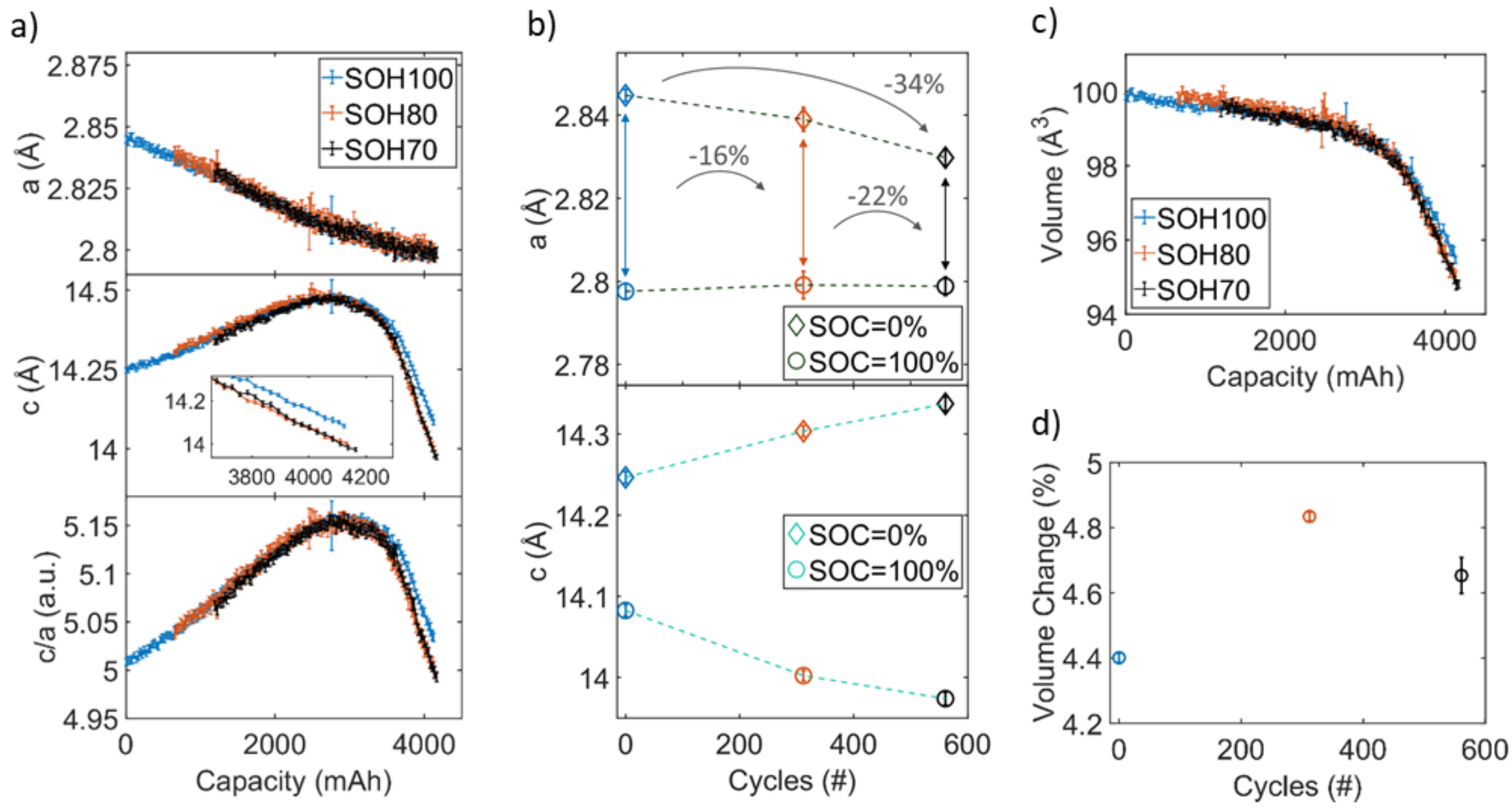

*Figure 4: (a): Change of the NCM unit cell parameters during CC charging. (b): Values of the lattice parameters at SOC=0% and SOC 100%. The relative change of the total movement of the a parameter is noted in the figure. (c): Change of the unit cell volume. (d): The total volume change of the three cells plotted against the cycle number. The illustrated errors result from the Rietveld refinement.*

## Lithiation of Graphite

On the anode side, only the graphite structure was analysed since Si becomes amorphous upon lithiation and, thus, is not detectable via diffraction. The structure of lithiated graphite intercalation compounds is characterized by multiple stages, defined by the number of carbon layers separating the lithium layers. [20,51] Graphite undergoes several phase transitions from $LiC_{72}$ to $LiC_{18}$ before the $LiC_{12}$ phase (stage 2) is reached, [52,53] which can be tracked by the steady shift of the graphite peaks (SI 4a). The fully lithiated $LiC_6$ phase is reached upon further lithiation, defined as stage 1 (SI 4b). In this study, the analysis foucses on the formation of $LiC_{12}$ and $LiC_6$ to assess the degradation modes of the anodes during cyclic aging.

In any case, the shift of the graphite 001 peak indicates the lithiation of graphite up to the $LiC_{12}$ phase formation (Figure 5a). Over cyclic aging, two phenomena are visible: On the one hand, the transition is shifted to higher potentials with lower SOH, which agrees with the shift in the capacity curves and is caused by the build-up of the internal resistance (Figure 2). On the other hand, less Li is required to induce the transition to $LiC_{12}$, as the SOH70 cell reaches the $LiC_{12}$ position faster when plotted against the capacity (Figure 5b). The slope of the shift in position from 43° to 41.5° is steeper for the more aged anodes. Since the starting position in the discharged state and the end position in the charged state are approximately similar, less Li is required for the transition, and the amount of active anode material decreases with continuous cycling. This agrees with the LAAM, determined in the DVA curves (Figure 3d). It was reported that loss of anode active material in graphite anodes is mainly caused by the mechanical breakdown during cycling [54–56] and exfoliation promoted by co-intercalation of manganese ions and solvent molecules [57]. Furthermore, a shoulder in the peak of the SOH70 cell is visible up to $\mathrm{SOC} \approx 40\%$, suggesting inhomogeneous lithiation of the anode (Figure 5c). For the SOH100 cell, this was not visible (SI 5).

Upon further lithiation, every second graphene slab is filled, and the $LiC_6$ phase (stage 1) is reached. Unlike the continuous C to $LiC_{12}$ transition, the formation of the $LiC_6$ phase is tracked by the increase of the $LiC_6$ reflection peak area at 39.4° (SI 4). Figure 5d shows the integrated intensity for the three cells during charging. The detected areas at low capacities are caused by falsely fitted peaks in the background signal and do not indicate any real $LiC_6$ formation. The onset of the $LiC_6$ phase formation begins at a smaller amount of transferred Li for the aged cells, which agrees with the beginning of the $LiC_{12}$ phase formation at smaller capacities with aging and suggests the LAAM. The onset capacity, which was determined by linear regression (SI 6), decreases almost linearly with the number of cycles (Figure 5e). The results agree with the trend of the MaxHi positions in the DVA curves, which also describe the transition from $LiC_{12}$ to $LiC_6$ (Figure 3c).

The $LiC_6$ peak area correlates to the amount of $LiC_6$ formed, which decreases with cycling and, thus, indicates that the amount of cyclable Li reduces. By comparing the $LiC_{12}$ and $LiC_6$ peak areas at $SOC = 100\%$, the loss of Li is illustrated (Figure 5f). The integrated $LiC_6$ intensity is the highest for the SOH100 cell and decreases with aging, while the intensity of the $LiC_{12}$ peak is the strongest for the SOH70 cell. This contrasting movement of the peak areas shows that not all available $LiC_{12}$ transitions to the fully lithiated $LiC_6$ phase for the aged cells, suggesting the loss of Li inventory (LLI). These results agree with the electrochemical data (Figure 3) and the analysis of the NCM cathode (Figure 4), where the loss of Li is also detected. For graphite-based anodes, the continuous SEI formation was identified as the primary reason for LLI. [58–60] Xie et al. have reported LLI as the leading degradation mode followed by loss of active anode material in 18650 cells with NCM|graphite electrodes, which agrees with our *in operando* ND results. [55]

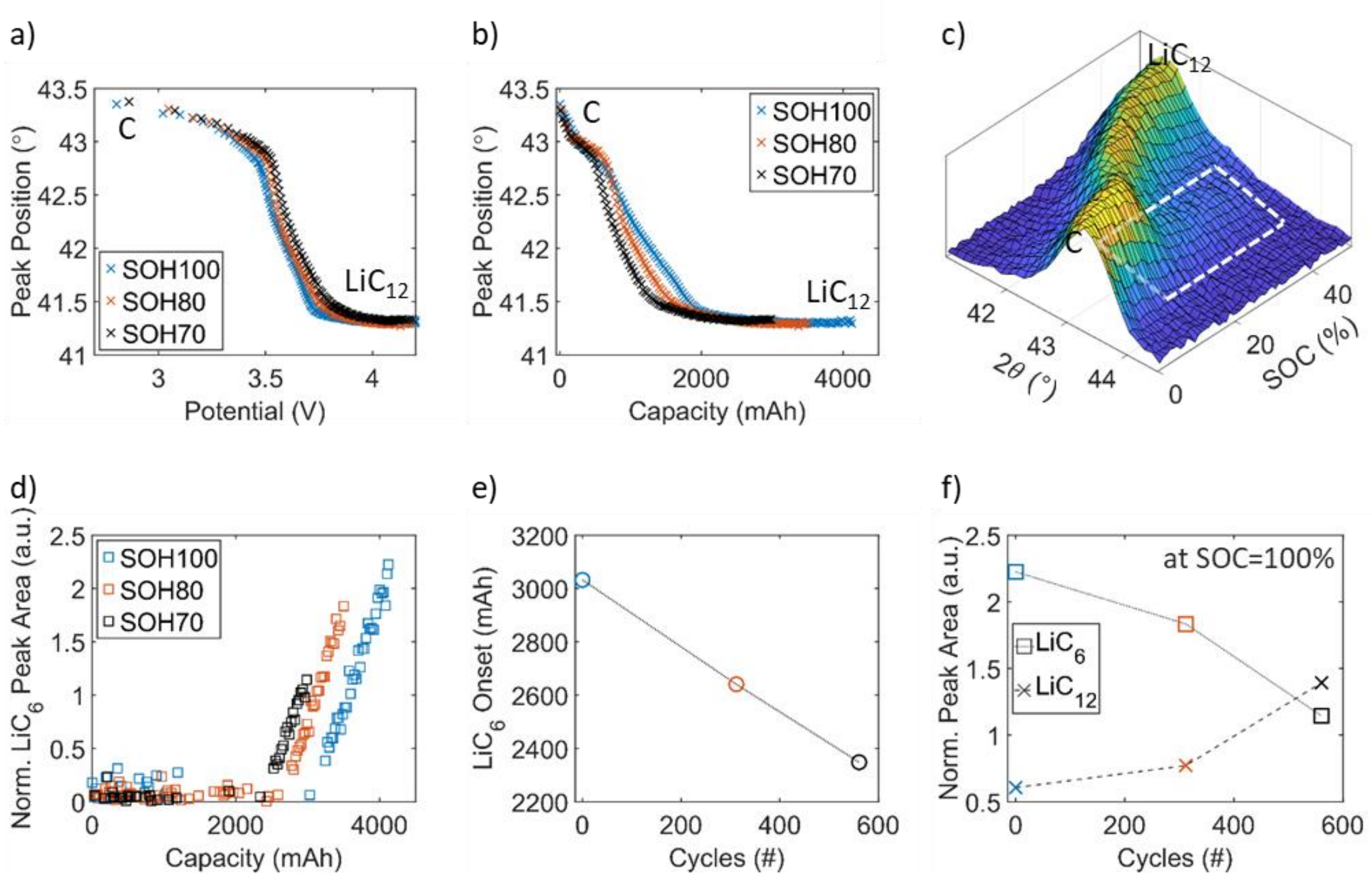

*Figure 5: Change of the graphite 002 peak position against the potential (a) and the capacity (b) during charging. 3D contour plot of the diffraction pattern for the SOH70 cell to highlight (white rectangle) the peak shoulder (c). Normalized $LiC_6$ 002 peak area against the capacity (d). The area of the $LiC_6$ and graphite peaks was normalized by the area of the NCM 003 peak in order to compare the different cells. Onset of the LiC6 formation against the cycle number (e). Comparison of the $LiC_6$ and C peak areas after different cycle numbes (f).*

## Lithium Concentration Profiles from NDP Measurements

From the *in operando* ND, Li loss is detected with cyclic aging, mainly missing on the cathode side. Therefore, NDP is performed to study the Li distribution in the electrodes at different SOH and at different positions (Figure 1c) to validate the assumption that the missing Li is consumed or trapped in the anode. Hereby, the $^3H$ particles with the highest energy are from regions close to the surface. The more material the $^3H$ particles have to penetrate, the more their energy will be reduced (SI 7). The Li concentration profiles for NCM at the discharged state are shown in Figure 6a & b. Generally, the Li is distributed very evenly across the thickness of the cathode, demonstrating a very uniform delithiation. At SOH100, the cathode has an overall higher Li concentration across the thickness of the cathode than at SOH70, confirming that Li is lost on the cathode side during cyclic aging. [add eigenes paper]

Comparing the two different cell positions, the cathodes at the inside edge (IE) have a slightly higher Li amount at both SOH than at the inside center (IC), indicating existing inhomogeneities along the height of the cylindrical cell. Petz et. al. have also reported the gradients inside 21700-type cells, where the Li amount in graphite fluctuated depending on the height. [8] Similarly, Bond et al. reported elevated $LiPF_6$ concentration in the middle compared to the top and bottom in 18650 containing C/SiOx and existing gradients in the different electrolyte solvents. [10]

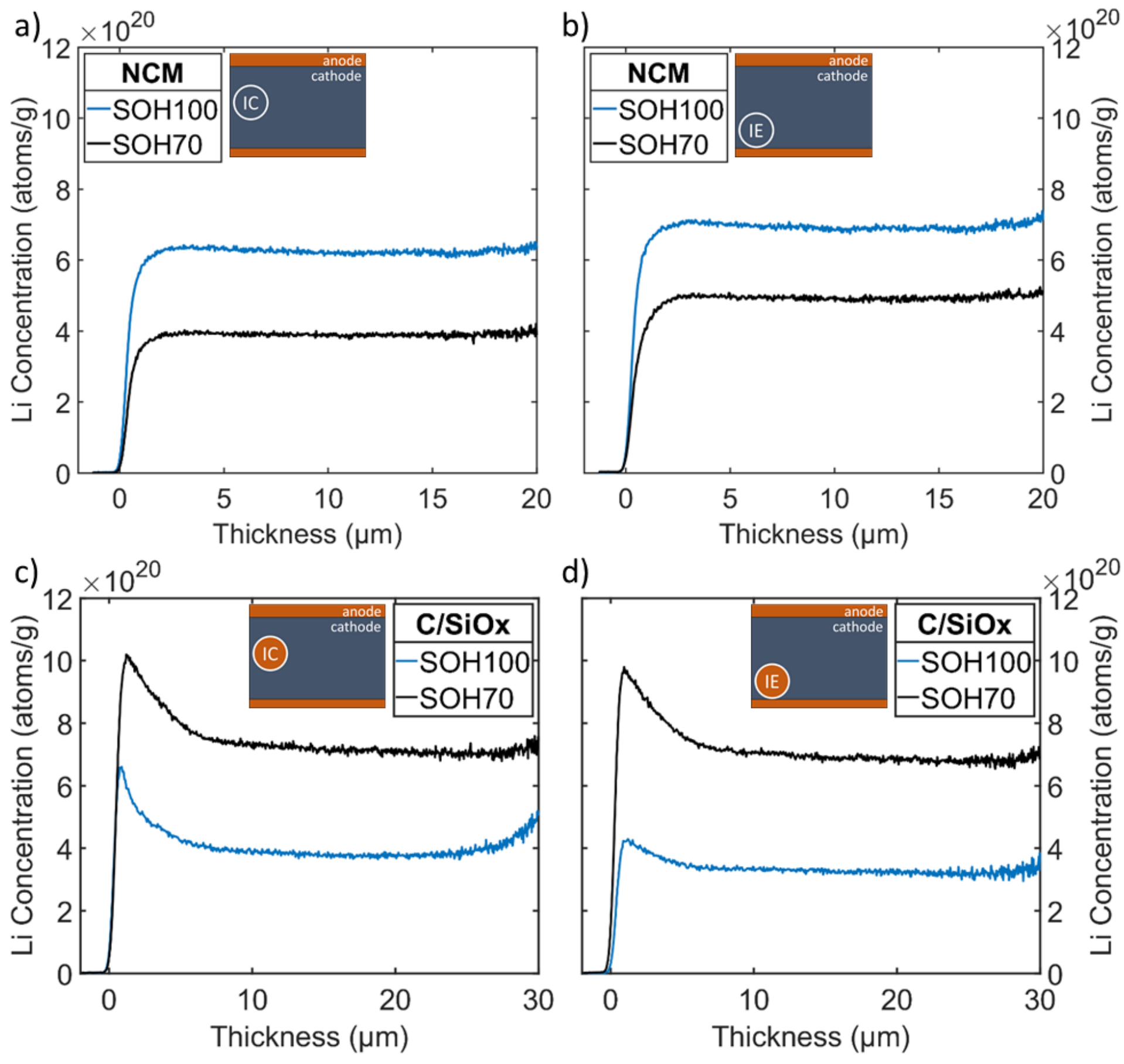

*Figure 6: Li concentration profiles for the NCM cathodes at the inside center (IC) (a) and inside edge (IE) positions (b) and the C/SiOx anodes at the IC (c) and IE (d) positions. For the calculation of the profiles, a uniform density was assumed for the electrodes.*

The Li depth profiles of the anodes are illustrated in Figure 6c & d. For all the anodes, the Li concentration curves have a sharp peak close to the surface and show an approximately linear decline towards the bulk of the anode, similar to the cathode. The bulk concentration is higher for the anode at SOH70 than at SOH100, which confirms that more Li is trapped in the anode with prolonged cycling. It should be noted that the SOH100 was cycled during the formation and the *in operando* ND experiment and, thus, is aged to a small extent, explaining the presence of Li in the anode.

A sharp surface peak is visible for all the samples, which most probably originates from the SEI. As the SEI is formed on the surface of the active anode material and contains Li, the $^3$H particles from the NDP reaction have the highest energy and, thus, can be assigned to Li originating from the surface. In previous NDP studies, similar surface peaks were observed on the anode surfaces, confirming the SEI formation. [60,61] Moving further towards the bulk, the signal from the SEI can either overlap with the Li inside the active material or the SEI of differently-sized particles. Therefore, the Li concentration is evened out, and no sharp peaks are detected. During cycling, cyclable Li is consumed by the continuous formation of the SEI, which was reported to grow with time [60,62,63] and is considered as the main cause for LLI. [55] This agrees with our finding that the surface peak of the SOH70 sample is broader and higher, suggesting a thicker SEI with prolonged cycling. Comparing the different cell positions shows that the anodes extracted from the bottom edge (IE) of the electrode have an overall lower concentration, demonstrating lower degradation. This is especially visible for the SOH70 anode, whose surface peak is significantly lower than for the SOH100 anode.

The total amount of Li inside the samples is calculated by determining the area beneath the concentration as shown in SI 8, which depends on the defined thickness of the electrodes that marks the end of the integration. Generally, the increased Li concentration with aging on the anodes is consistent with the decreased Li concentration on the cathodes (Figure 7). For the NCM at IC position, the Li concentration decreases by 38% and exceeds the loss that is detected at the IE position. Generally, these values are in agreement with the reduction of the a parameter movement by 34% that was detected with the *in operando* ND. This finding shows that the loss of Li inside the NCM can be successfully determined by the change of the lattice parameters over cycling. The deviation can be explained by the small sample volume of the NDP measurement in comparison to the ND, as the ND results represent the average of the probed area. The Li concentration in the anodes increases by a lot during cycling, emphasizing that the capacity loss is mainly caused by the anode. Here, the difference between the positions is not as large as for the cathodes. In theory, the summation of the Li concentration of both electrodes should yield similar results. However, the total amount of Li is higher after cyclic aging, suggesting that more Li is in the electrodes than in the uncycled state. The additional Li might originate from the Li salt in the electrode. Generally, small deviations due to the small sample volume compared to the large cell format cannot be excluded.

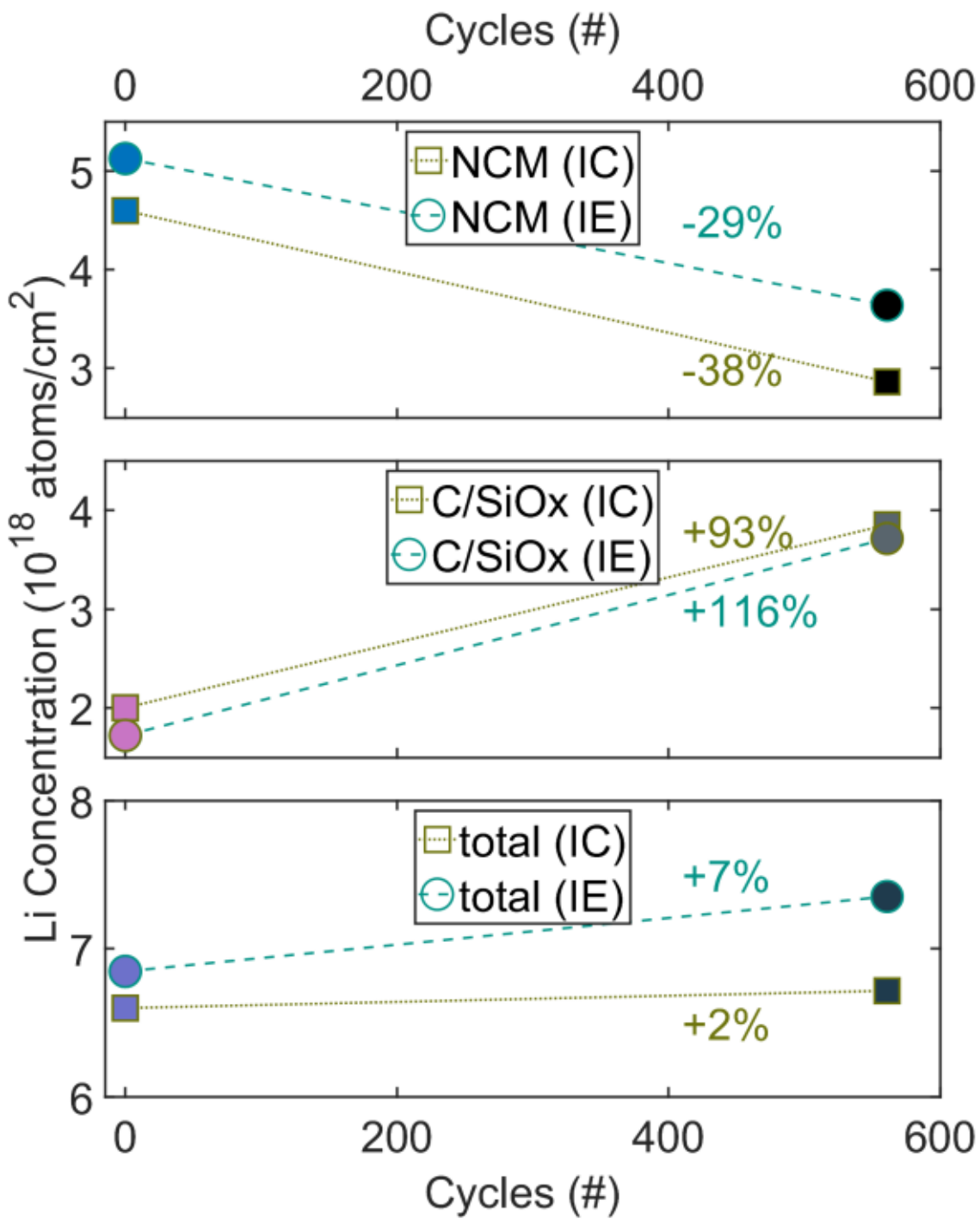

*Figure 7: The total Li concentration of the cathodes (top), anodes (middle) at the inside center (IC) and inside edge (IC) positions and the sum of them (bottom) for the SOH100 and SOH70 cells. The relative change from SOH100 and SOH70 is noted inside the figure. The dashed and dotted lines function as a visual aid.*

## Conclusions

This study investigated the influence of cyclic aging on NCM cathodes and C/$SiO_x$ anodes in 21700-type cells and identified loss of Li and loss of active anode material as reasons for the capacity loss. During cycling, the decrease in the discharge capacity is detected, confirming the loss of capacity. From the incremental capacity analysis (ICA) and differential voltage analysis (DVA), loss of Li inventory (LLI) and loss of active anode material (LAAM) were detected. In the DVA, the degradation of the $SiO_x$ was visible, which contributes strongly to the pronounced LAAM. *In operando* neutron diffraction (ND) is performed to study the structural changes of the electrodes during cycling. With a higher aging degree, two effects could be identified for the NCM cathodes: First, the lithiation degree of the NCM in the discharged state is reduced with aging, indicating the amount of Li returning from the anode during discharge is decreased. Second, the NCM is delithiated to higher degrees with aging, which is reflected in the lattice parameters and suggests that the cathode potentials are shifted to higher values. The increased delithiation of the NCM leads to a more significant volume change of the active materials that induce stress and facilitate particle cracking. Furthermore, the aging on the anode side is investigated by analyzing the lithiated graphite phases. With aging, the amount of the fully lithiated $LiC_6$ phase decreases while the amount of $LiC_{12}$ increases, reflecting LLI. Additionally, the amount of Li necessary for reaching the $LiC_{12}$ and $LiC_6$ phases decreases for the aged cells, demonstrating extensive LAAM and confirming the DVA results. The ND results suggest that the capacity loss mainly originates from the anode side. These findings are in agreement with the NDP measurements of the extracted electrodes. The LLI is identified in the cathode as the cathode at SOH100 contains more Li than at SOH70. Simultaneously, the Li concentration in the anodes at SOH70 increases significantly, which underlines that Li loss primarily originates from the anode side. The SEI growth with cyclic aging is detected by the surface peaks in the Li concentration, which are broader and higher for the SOH70 anodes. Comparing the Li depth profiles at the center and the bottom edge of the 21700-type cells shows that the degradation effects are higher at the center position and stresses the inhomogeneous aging. In the future, analyzing the cell across all heights can give deeper insight into cell degradation. It might improve the cell design in order to minimize inhomogeneities and, thus, enhance the battery performance.

## Acknowledgements

This work has received funding from the German Federal Ministry of Education and Research (BMBF) under the grant 'ExZellTUM III' (03XP0255) within the ExcellBattMat cluster, from the German Federal Ministry for Economic Affairs and Climate (BMWK) under the grant 'CAESAR' (03EI3046F) and by the Technical University of Munich (TUM). It was also supported by the Ministry of Education, Youth and Sports (MEYS) CR under the project OP JAK CZ.02.01.01/00/22_008/0004591. The authors thank BASF SE and VARTA AG (N. Bucher, S. Schebesta) for providing the materials used in this study. This work is partly based on experiments carried out at the Swiss Spallation Neutron Source SINQ, Paul Scherrer Institute, Villigen, Switzerland (proposal 20230258) and at the CANAM infrastructure of the NPI CAS Rez, Czech Republic.

## Supporting Information

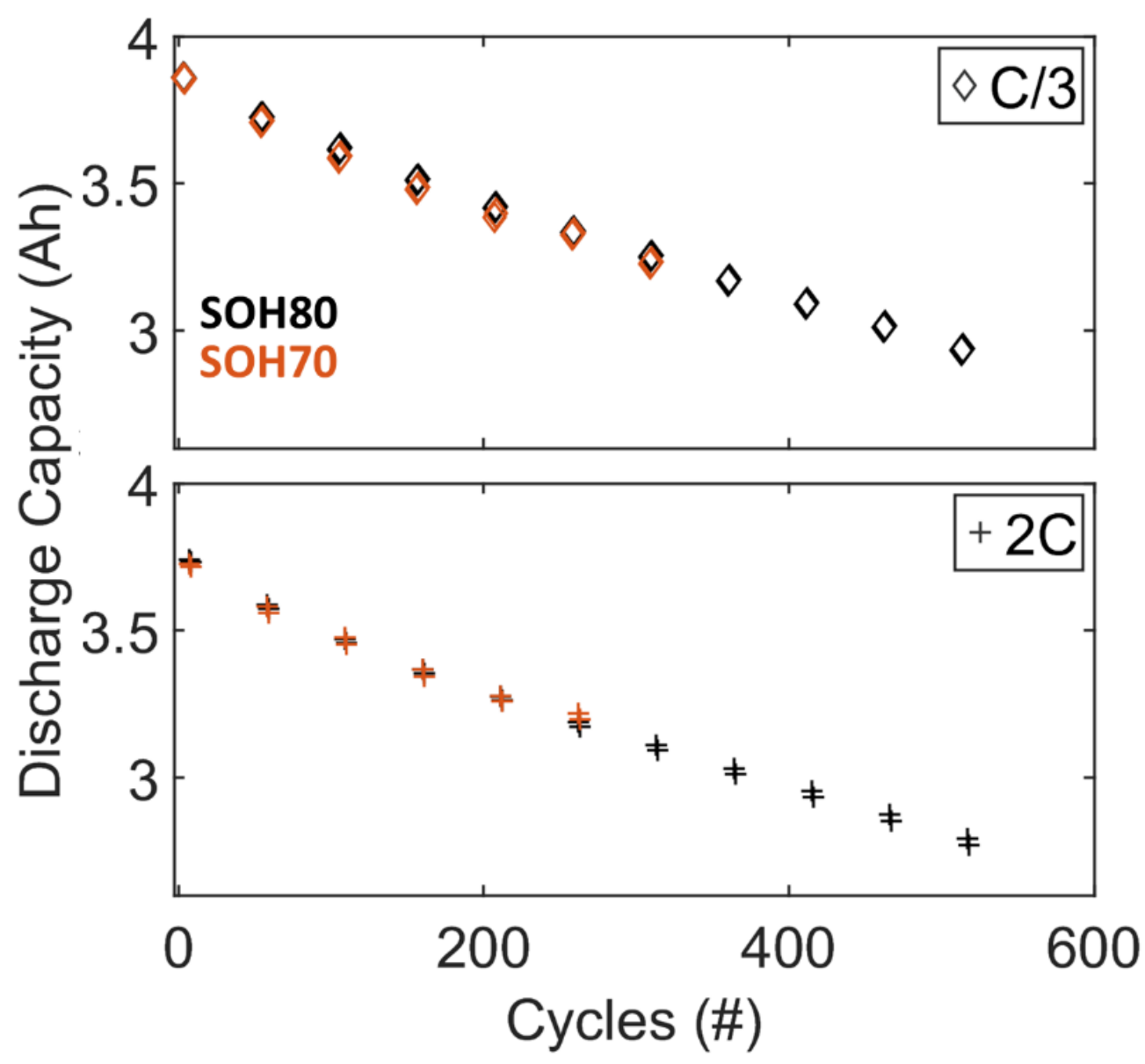

*SI 1: Discharge capacities for cylces at C/3 and 2C over cycling.*

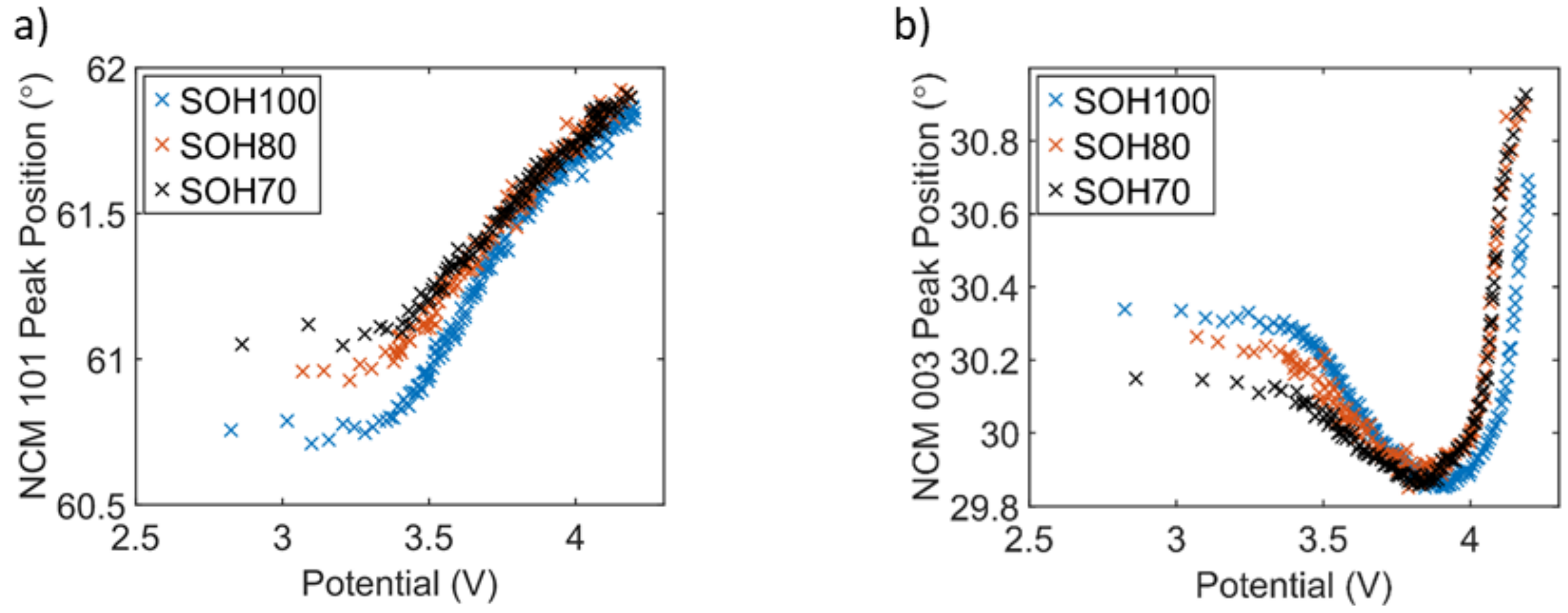

*SI 2: Evolution of the NCM 101 (a) and NCM 003 (b) peak positions during the charge step for the three cells.*

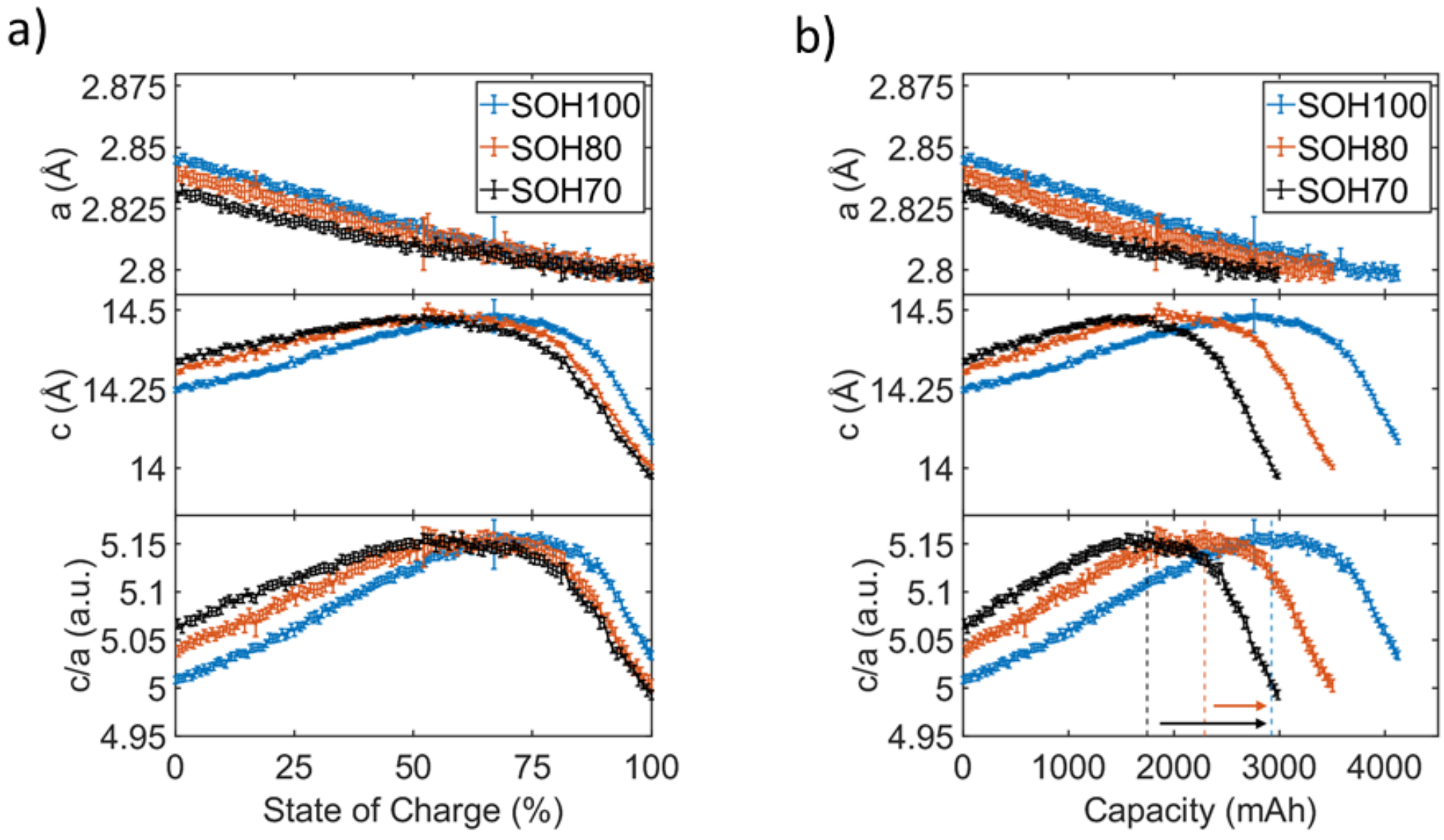

*SI 3: Evolution of the NCM lattice parameters against the state of charge (a) and the transferred capacity (b). For the analysis, the peak in the c/a curve was determined and the capacity for the SOH70 and SOH80 cells were shifted until the peaks overlap.*

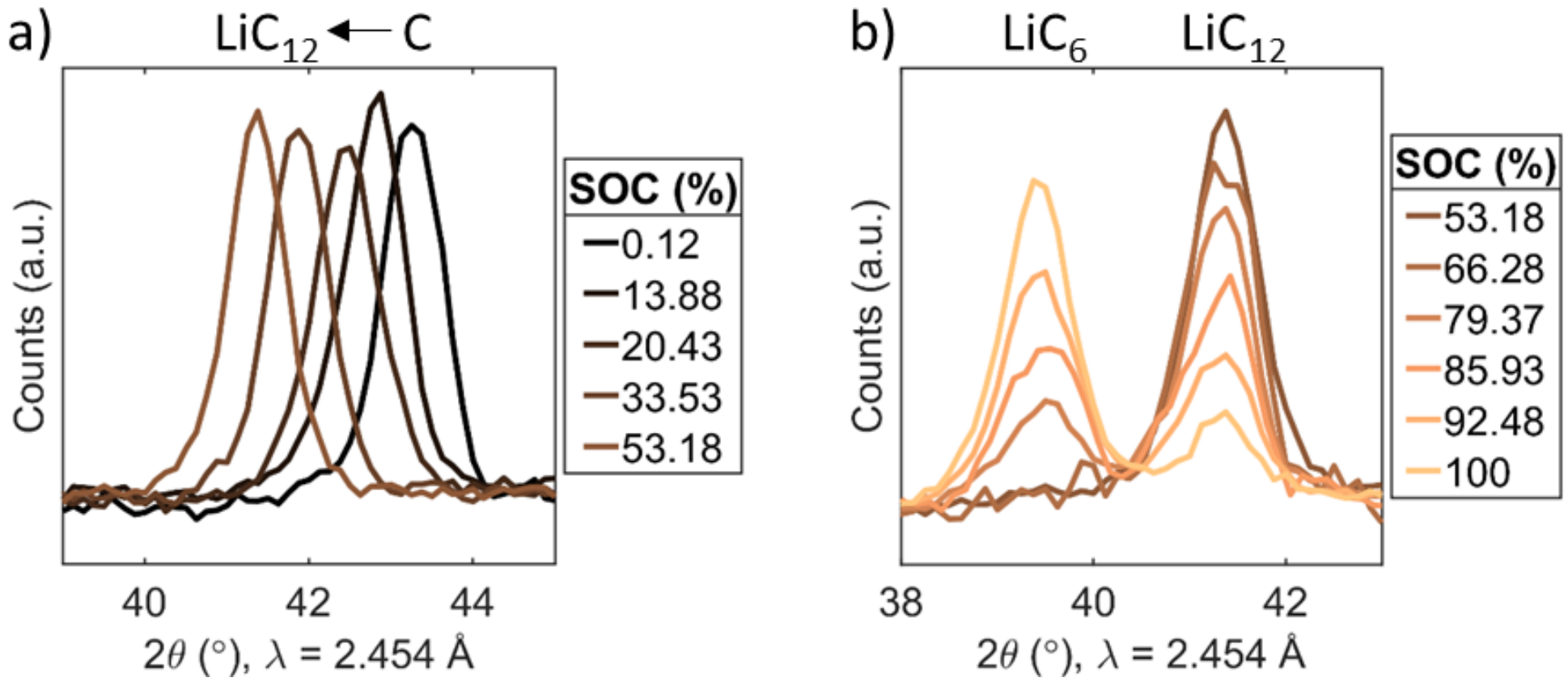

*SI 4: Segments of the diffraction patterns of the SOH100 cell that show the lithiation of graphite up to $LiC_{12}$ (a) and the transition from $LiC_{12}$ to $LiC_6$ (b).*

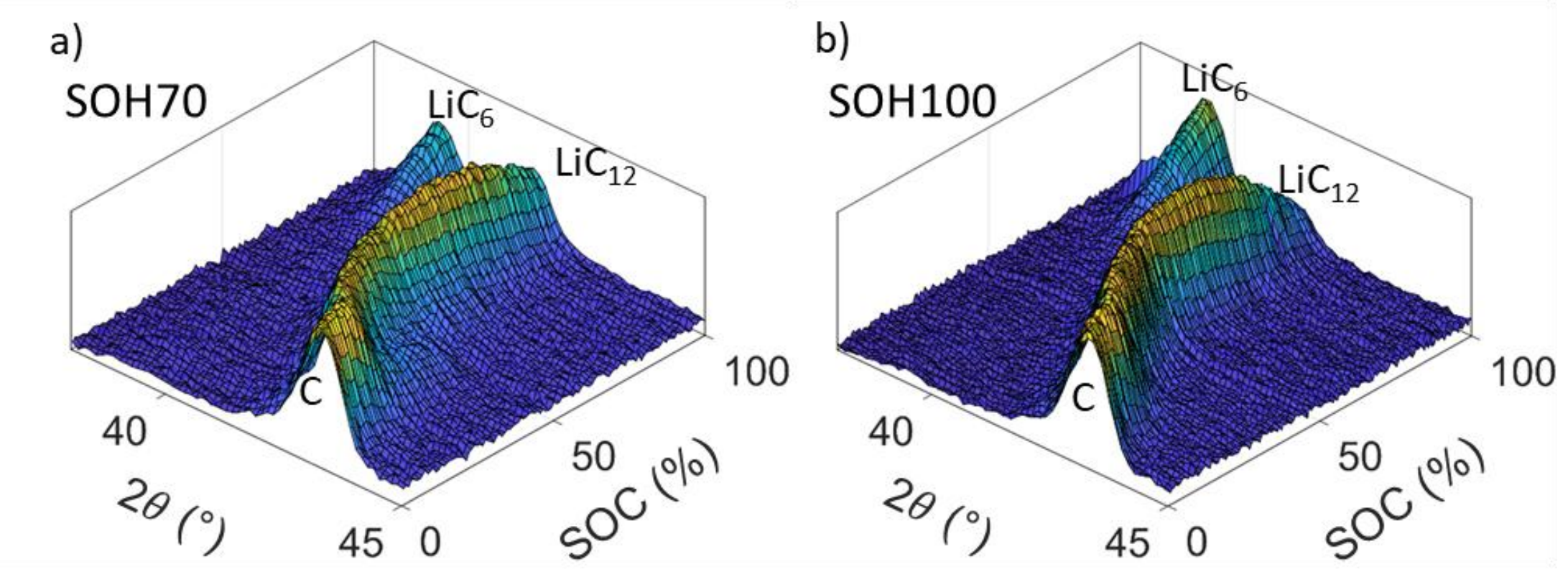

*SI 5: 3D contour plot of the diffraction patterns for the SOH70 (a) and SOH100 (b) cell in the $2\theta$ region of the transition from graphite to $LiC_{12}$ and $LiC_6$ to demonstrate the inhomogeneous phase transition in the SOH70 cell. In contrary to the SOH100 cell, the existence of two peaks can be seen at low SOC for the SOH70 cell .*

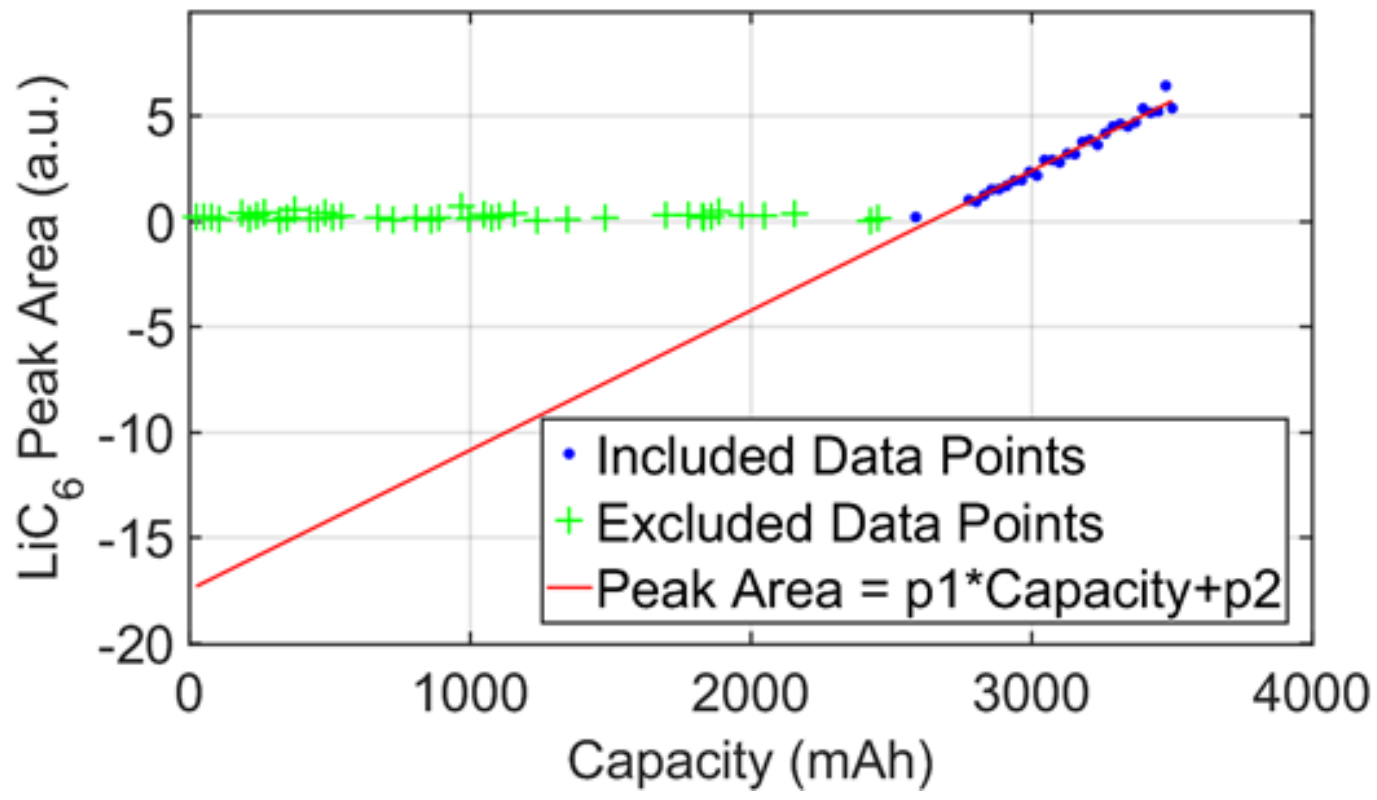

*SI 6: Determination of the onset of the $LiC_6$ formation by performing a linear fit of the peak area evolution.*

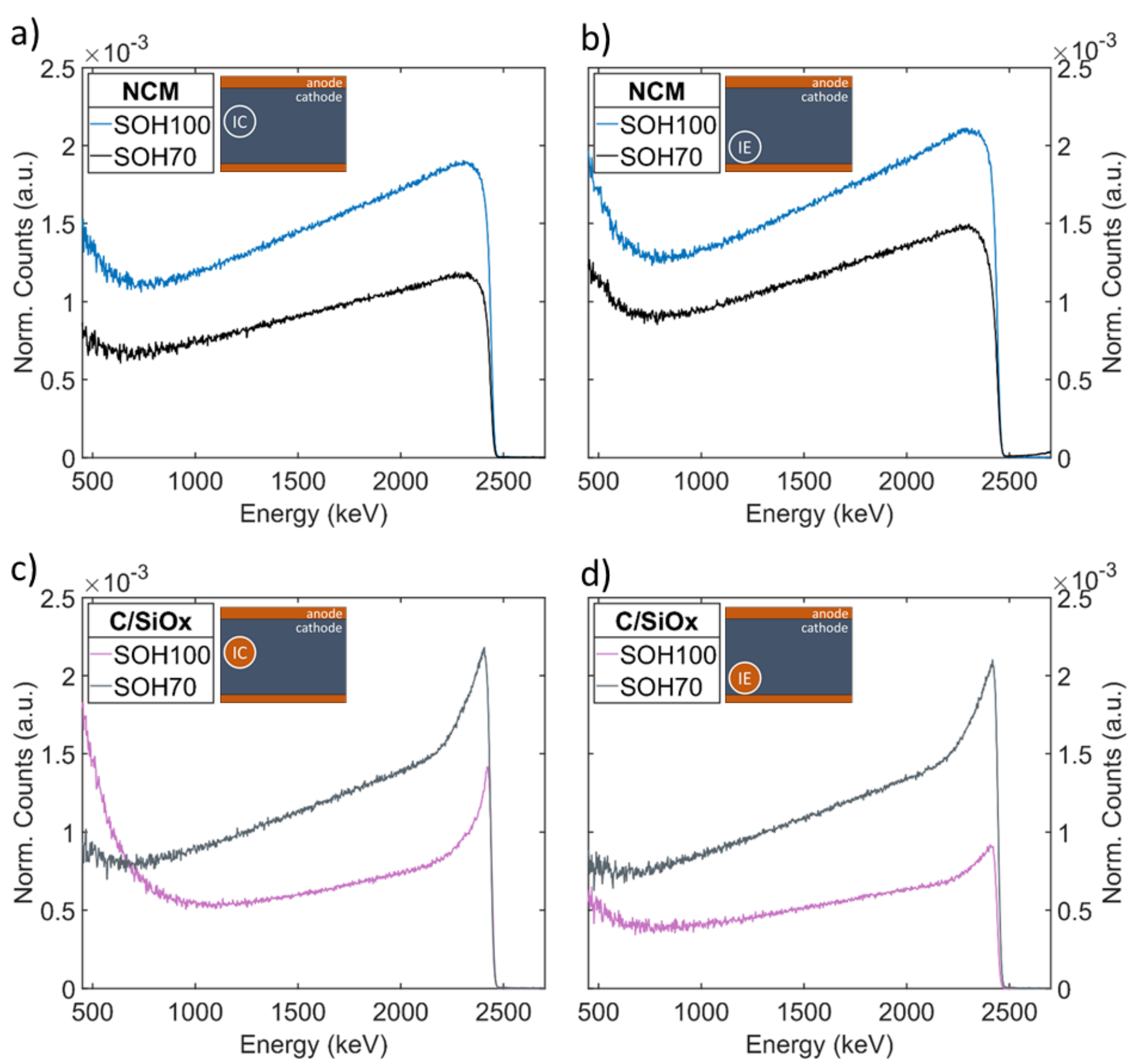

*SI 7: Normalized counts plotted against the energy of the emitted particles for the cathode at the inside center (IC) (a) and inside edge (IE) position (b) and the anode (b, c). The counts were normalized by the neutron fluence during the experiment.*

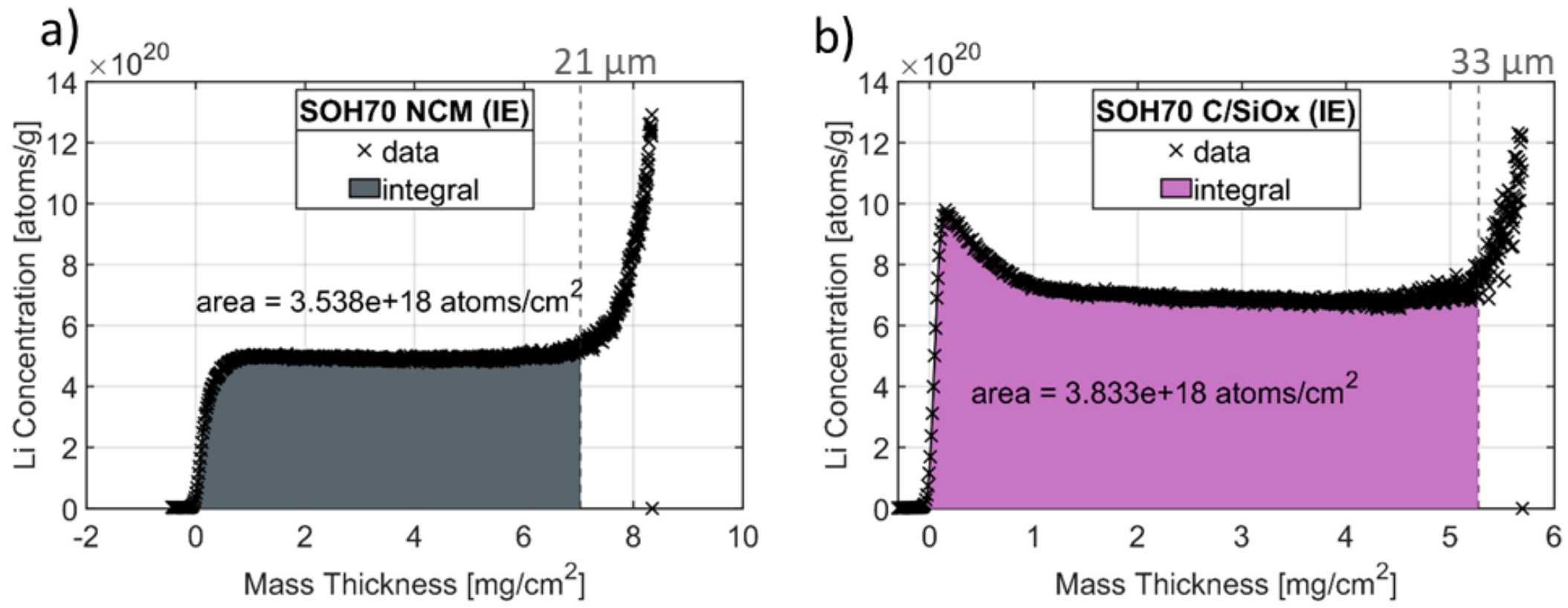

*SI 8: Integration of the Li profiles to determine the total amount of Li inside the cathode (a) and anode (b).*